\documentclass[aps,prl,reprint,amsmath,superscriptaddress]{revtex4-1}
\usepackage{graphicx}
\usepackage{hyperref}

\begin{document}
\title{Controlling Chirality of Entropic Crystals}

\author{Pablo F. Damasceno}
\email{damascus@umich.edu}
\affiliation{Applied Physics Program, University of Michigan, Ann Arbor, Michigan 48109, USA}
\affiliation{Biointerfaces Institute, University of Michigan, Ann Arbor, Michigan 48109, USA}

\author{Andrew S. Karas}
\affiliation{Biointerfaces Institute, University of Michigan, Ann Arbor, Michigan 48109, USA}
\affiliation{Department of Chemical Engineering, University of Michigan, Ann Arbor, Michigan 48109, USA}

\author{Benjamin A. Schultz}
\affiliation{Biointerfaces Institute, University of Michigan, Ann Arbor, Michigan 48109, USA}
\affiliation{Department of Physics, University of Michigan, Ann Arbor, Michigan 48109, USA}

\author{Michael Engel} 
\affiliation{Biointerfaces Institute, University of Michigan, Ann Arbor, Michigan 48109, USA}
\affiliation{Department of Chemical Engineering, University of Michigan, Ann Arbor, Michigan 48109, USA}

\author{Sharon C. Glotzer}
\email{sglotzer@umich.edu}
\affiliation{Applied Physics Program, University of Michigan, Ann Arbor, Michigan 48109, USA}
\affiliation{Biointerfaces Institute, University of Michigan, Ann Arbor, Michigan 48109, USA}
\affiliation{Department of Chemical Engineering, University of Michigan, Ann Arbor, Michigan 48109, USA}
\affiliation{Department of Physics, University of Michigan, Ann Arbor, Michigan 48109, USA}
\affiliation{Department of Materials Science and Engineering, University of Michigan, Ann Arbor, Michigan 48109, USA}

\date{\today}

\begin{abstract}

Colloidal crystal structures with complexity and diversity rivaling atomic and molecular crystals have been predicted and obtained for hard particles by entropy maximization.
However, so far homochiral colloidal crystals, which are candidates for photonic metamaterials, are absent.
Using Monte Carlo simulations we show that chiral polyhedra exhibiting weak directional entropic forces self-assemble either an achiral crystal or a chiral crystal with limited control over the crystal handedness.
Building blocks with stronger faceting exhibit higher selectivity and assemble a chiral crystal with handedness uniquely determined by the particle chirality.
Tuning the strength of directional entropic forces by means of particle rounding or the use of depletants allows for reconfiguration between achiral and homochiral crystals.
We rationalize our findings by quantifying the chirality strength of each particle, both from particle geometry and potential of mean force and torque diagrams.

\end{abstract}
\pacs{asa}
\maketitle

Controlling --and understanding the origins of-- chirality is a major goal in the physical and chemical sciences.
It has longstanding implications that range from the design of effective pharmaceuticals to explaining biological homochirality and the beginnings of life~\cite{bernal1967origin}.
In recent years the expectation that chiral materials can provide a route for photonic metamaterials~\cite{Pendry2004} has renewed the interest in chirality, now from a materials design perspective.
A key remaining challenge for the development of metamaterials~\cite{Soukoulis2011} is the ability to use scalable techniques to create crystals with unique chirality.
Of particular importance are enantioselective processes, which select a pre-defined handedness of the crystal, thereby maximizing optical activity.

While several routes have been reported to assemble structures with chiral order, from exploitation of electrostatic~\cite{Vernizzi2009}, anisotropic~\cite{Fejer2010} and isotropic~\cite{Edlund2012} interactions, packing~\cite{Mughal2011,Zhao2012,Gantapara2015}, spontaneous symmetry breaking in twisted ribbons~\cite{Srivastava2010,Singh2014b}, to self-sorting in liquids~\cite{Dressel2014}, the lack of chiral forces in these examples preclude homochirality.
Even if chiral forces such as circularly polarized light~\cite{Yeom2015} are present, or when assembling naturally handed viruses~\cite{Gibaud2012}, bacteria~\cite{Barry2006} or nanoparticles with pre-sorted chiral coatings ~\cite{Ben-Moshe2014}, or even by judiciously designing chiral building blocks ~\cite{Schreiber2013,Rossi2015,Dussi2015,Tian2015}, the final chiral structures have, so far, been restricted to liquid crystalline and low-dimensional arrangements.
Growth of three-dimensional homochiral colloidal crystals has yet to be demonstrated.
In fact, the only examples of chiral crystals with realized applications as polarization sensitive devices~\cite{VonFreymann2010a,Turner2013} were fabricated via non-scalable processes such as direct laser writing, hampering large scale production.

\begin{figure}
\centering
\includegraphics[width=\columnwidth]{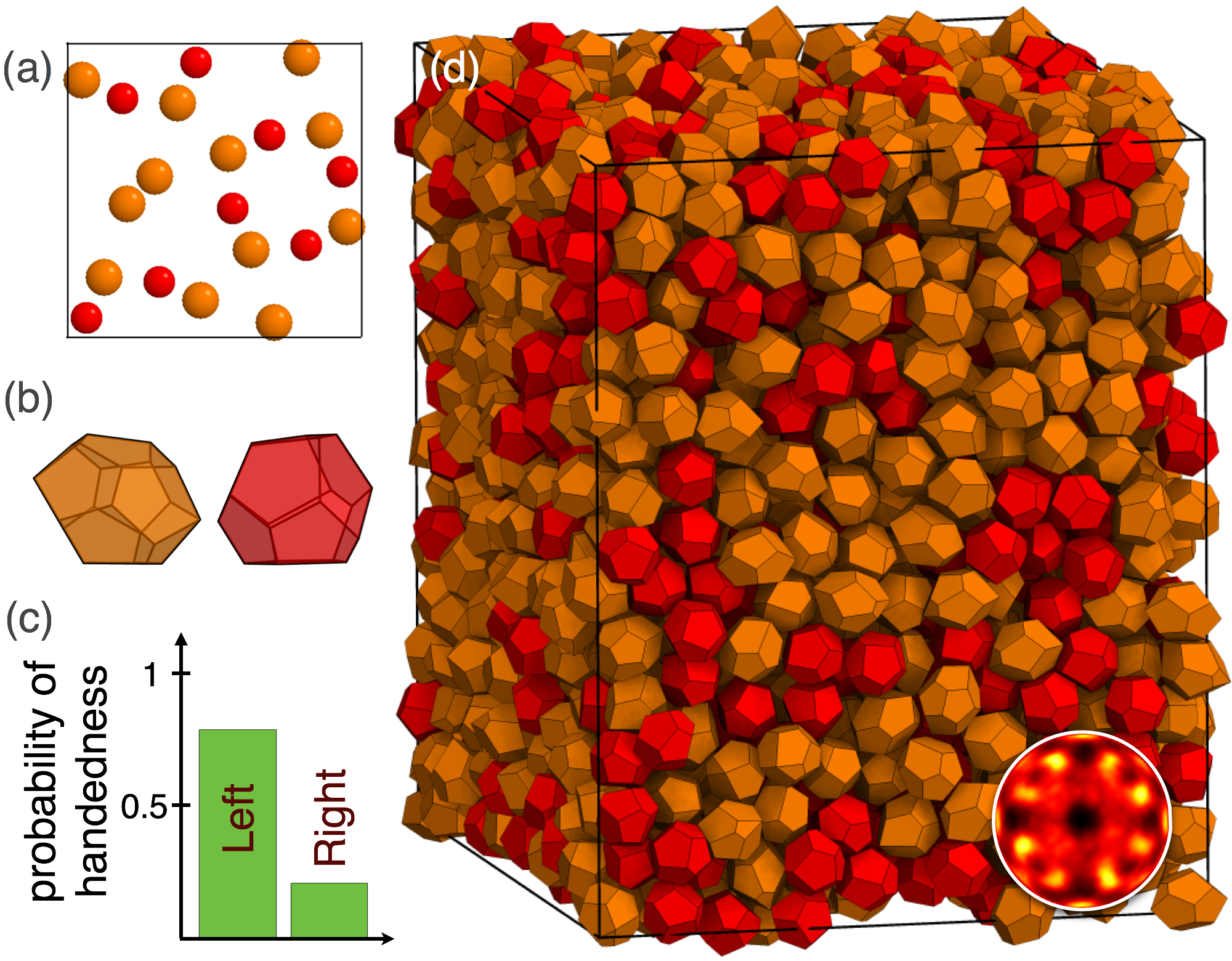}
\caption{
Assembly of the $\beta$-Mn crystal structure from its VPs.
(a)~The unit cell with 20 particles.
(b)~The VP shown on the left (orange) is elongated with 14 faces and 22 vertices.
The VP shown on the right (red) resembles a dodecahedron with 12 irregular pentagonal faces and 20 vertices.
(c)~We observe the assembly of both left-handed and right-handed variants, with a bias for left-handed, the handedness dictated by the VPs.
(d)~A snapshot of the assembled crystal shown in projection.
The bond orientational order diagram (inset) allows identifying the crystal structure.
The crystal is degenerate, which means either of the two VPs can be found at a given lattice point with comparable probability.
}
\label{fig1}
\end{figure}

Recent work has shown that hard polyhedra can self-assemble into a great diversity of crystals, liquid crystals, and quasicrystals via entropy maximization~\cite{Frenkel1999,Haji-Akbari2009,Damasceno2011,Agarwal2011,Damasceno2012,Henzie2012}, a mechanism that can be understood as resulting from emergent directional entropic forces (DEFs) between neighboring particles due to crowding~\cite{Damasceno2011,vanAnders2014,vanAnders2015}.
Even chiral crystals have been assembled.
The $\beta$-Mn crystal structure with chiral space group P4$_1$32 [Fig.~\ref{fig1}a], earlier reported in spheres interacting via an isotropic oscillatory pair potential~\cite{Elenius2009}, self-assembles from achiral dodecahedra shapes~\cite{Damasceno2012}.
However, as expected, the handedness of the crystal could not be controlled and both left- and right-handed crystals were observed with equal probability.
Given the ability of entropic interactions to imitate the plethora of complex structures achievable from atomistic interactions, the absence of homochiral entropic crystals is both practically discomforting and theoretically puzzling.

In this Letter, we investigate whether entropic forces alone can be sufficient for the growth of homochiral crystals.
By using a recently developed approach for shape design~\cite{Schultz2014}, we demonstrate that chiral building blocks can indeed grow three-dimensional chiral crystals of chosen handedness provided a moderate chiral entropic force between neighboring particles can be achieved.
We also use this method to show how chiral particles can be reversibly reconfigured to enforce a homochiral crystal when the entropic force strength is tuned by either shape `sharpening' (or unrounding) or by an increase in depletion concentration.
Our results can be rationalized in terms of features in particle geometry and suggest design strategies currently accessible to laboratory techniques that can be used to assemble homochiral crystals.

\textit{Voronoi particle approach.}---The forces controlling the self-assembly of non spherical hard particles are emergent and become more directional as particles become more aspherical~\cite{Damasceno2012,Damasceno2011,Young2013}.
In fact, by identifying shapes capable of maximizing DEFs in a specific crystal, entropically patchy particles can be inversely designed~\cite{vanAnders2014,vanAnders2014b,vanAnders2015}.
One approach is to construct Voronoi particles (VPs)~\cite{Schultz2014}.
VPs are polyhedra whose shapes are identical to the Voronoi cell of the target crystal and whose assembly at infinite pressure is, by construction and in the absence of degeneracy, guaranteed thermodynamically to be the target structure.

We apply the VP approach to the case of $\beta$-Mn in an attempt to select the handedness of the crystal.
From the perfect crystal structure [Fig.~\ref{fig1}a] we construct its two geometrically distinct chiral VPs [Fig.~\ref{fig1}b], one for each set of Wyckoff positions.
Although our prior work predicts that particles with high sphericity as measured by their isoperimetric quotient (IQ) self-assemble into a rotator crystal with weak face-to-face contacts between neighboring particles~\cite{Damasceno2012}, $\beta$-Mn VPs have IQ = (0.742, 0.756), values at the boundaries between crystals and rotator crystals, such that no clear prediction can be made.
We employ Monte Carlo simulations as in~\cite{Damasceno2011} to self-assemble the VPs from a disordered fluid state.
Systems with 2048 particles were equilibrated in the $NVT$ ensemble at packing fractions in the range $0.5\leq\phi\leq0.7$.
Previous studies of hard particle self-assembly suggest this range to be sufficiently high to observe crystallization while avoiding kinetic arrest.
Structure identification was conducted by calculating diffraction patterns and bond order diagrams~\cite{Damasceno2012}.

\begin{figure}
\centering
\includegraphics[width=\columnwidth]{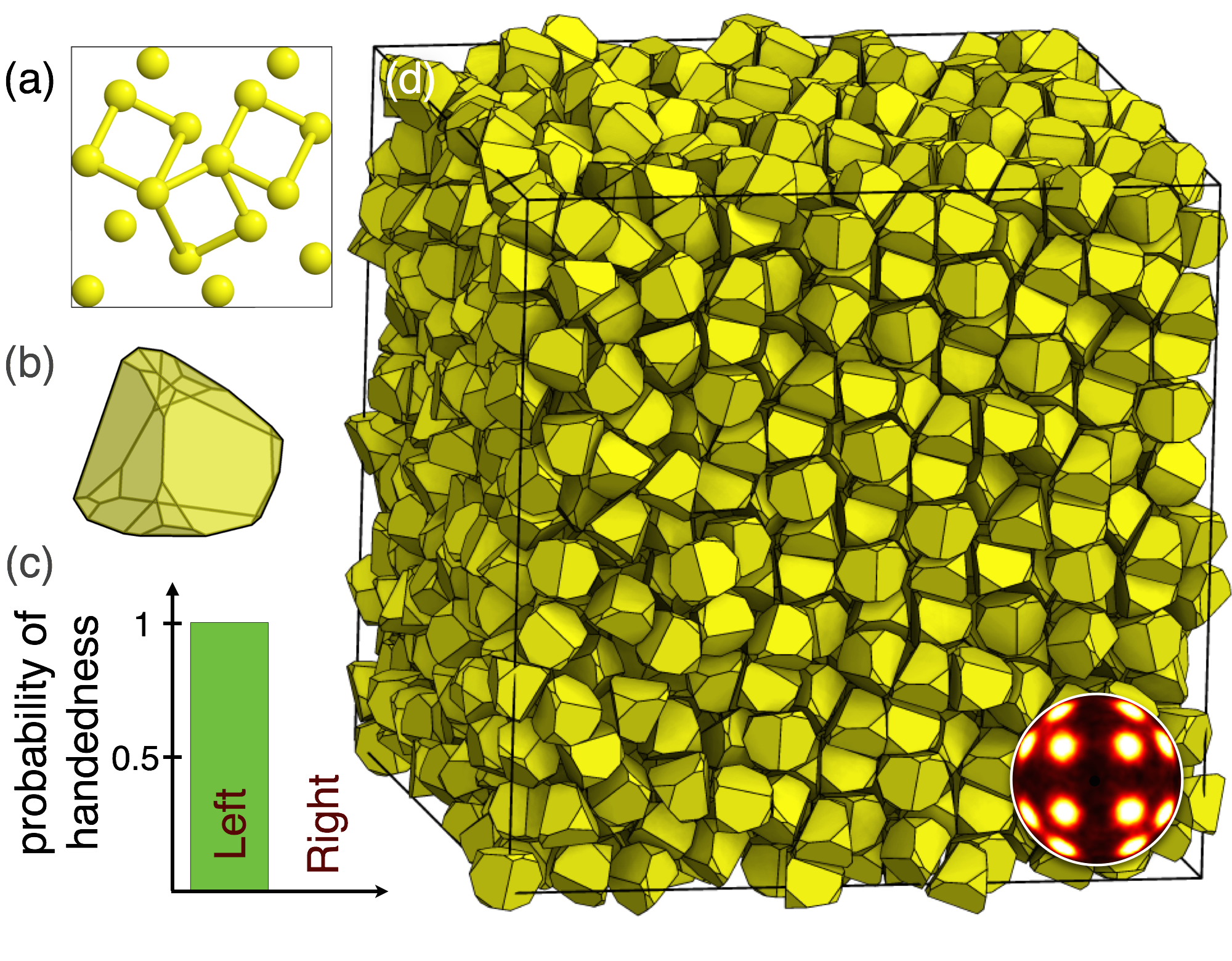}
\caption{
Assembly of the simple chiral cubic (SCC) crystal structure from its VPs.
(a)~A 2x2 patch of the unit cell with 4 particles.
The bonds indicate chiral helices with pitch length four.
(b)~The VP has point group $D_3$ and six symmetry-equivalent big faces.
(c)~The assembled crystal is homochiral with handedness determined by the building block.
(d)~A snapshot of the assembled crystal structure shown in projection.
}
\label{fig2}
\end{figure}

For packing fractions $\phi\geq0.58$ we observe that the system crystallizes into the target crystal. 
Just like in systems of hard dodecahedra~\cite{Damasceno2012} and Zetterling spheres~\cite{Zetterling2003}, we also notice the assembly of $\gamma$-brass and body-centered cubic (BCC) phases competing with the formation of $\beta$-Mn. 
In fact, out of 80 simulations run at these packing fractions, only 9 were found to crystallize into $\beta$-Mn, out of which 7 assembled into a crystal with the same handedness as the building blocks [Fig.~\ref{fig1}c]. 
Despite this statistical handshake between building block and bulk crystal chirality, the observation of crystals with unexpected chirality suggests that the overall handedness cannot be guaranteed \textit{a priori} and a stronger face-to-face interaction is needed.

It is worth noticing that we also observe occupational degeneracy between the two different VPs [Fig.~\ref{fig1}d].
This observation reveals that despite the significant geometric difference between VPs, their free-energy incentive to remain in their respective Wyckoff positions does not surpass the gain in mixing entropy~\cite{Haji-Akbari2011, Schultz2014}.
To understand this result, we refer back to the work of Schultz et al.~\cite{Schultz2014} where it was demonstrated that the Voronoi approach is generally successful if two conditions are satisfied:
(i)~The unit cell of the target crystal is small, and (ii)~the VPs are sufficiently symmetric and of only intermediate sphericity to generate sufficiently strong DEFs.
If these prerequisites are not satisfied, rotator phases and degenerate crystals become possible, as found here.
This suggests the sphericity of the $\beta$-Mn VPs might be too high to warrant pronounced face-to-face contact, and the chiral selectivity is not strong enough for enantioselective growth.

To improve the assembly probability of a chiral crystal and to obtain a homochiral crystal, we then hypothesize that a more facetted polyhedron, \textit{i.e.\ }one leading to stronger DEFs, is needed to imprint the chirality of the building block onto the assembled structure.
As a test, we consider a simpler target crystal than $\beta$-Mn.
In a recent investigation~\cite{Engel2014} it was discovered that a system of spheres interacting via a simple isotropic oscillatory pair-potential assembles a cubic chiral crystal with Pearson symbol cP4 and the same chiral space group P4$_{\text{1}}$32 as $\beta$-Mn.
Interestingly, a search through the International Tables of Crystallography~\cite{Hahn2006} reveals it to be the only crystal up to symmetry equivalence with a chiral cubic space group and four or less particles in the unit cell.
We therefore refer to this crystal as the simple chiral cubic (SCC) crystal.
Although proposed theoretically in high-pressure lithium~\cite{Ma2008} and semiregular nets~\cite{DelgadoFriedrichs2003}, SCC has not been self-assembled experimentally in natural or artificial systems.
Its low coordination number \text{CN} = 6 together with the low sphericity IQ = 0.564 of its VP [Fig.~\ref{fig2}b] makes SCC a good candidate for the entropy-driven assembly of a homochiral crystal.
Indeed, we observe the robust self-assembly of chiral crystals with a unique handedness [Fig.~\ref{fig2}c] in simulations [Fig.~\ref{fig2}d] at packing fractions $\phi\geq0.60$.
Between these simulations and those of rounded shapes (see below), a total of 92 simulations assembled into the SCC crystal, all of them with handedness dictated by the VP.
This result confirms that large faceting and low coordination number can lead to handedness selection in chiral crystals.
A quantitative analysis of the strength of DEFs required for such biasing will be investigated in the last section of this letter.

\begin{figure}
\centering
\includegraphics[width=\columnwidth]{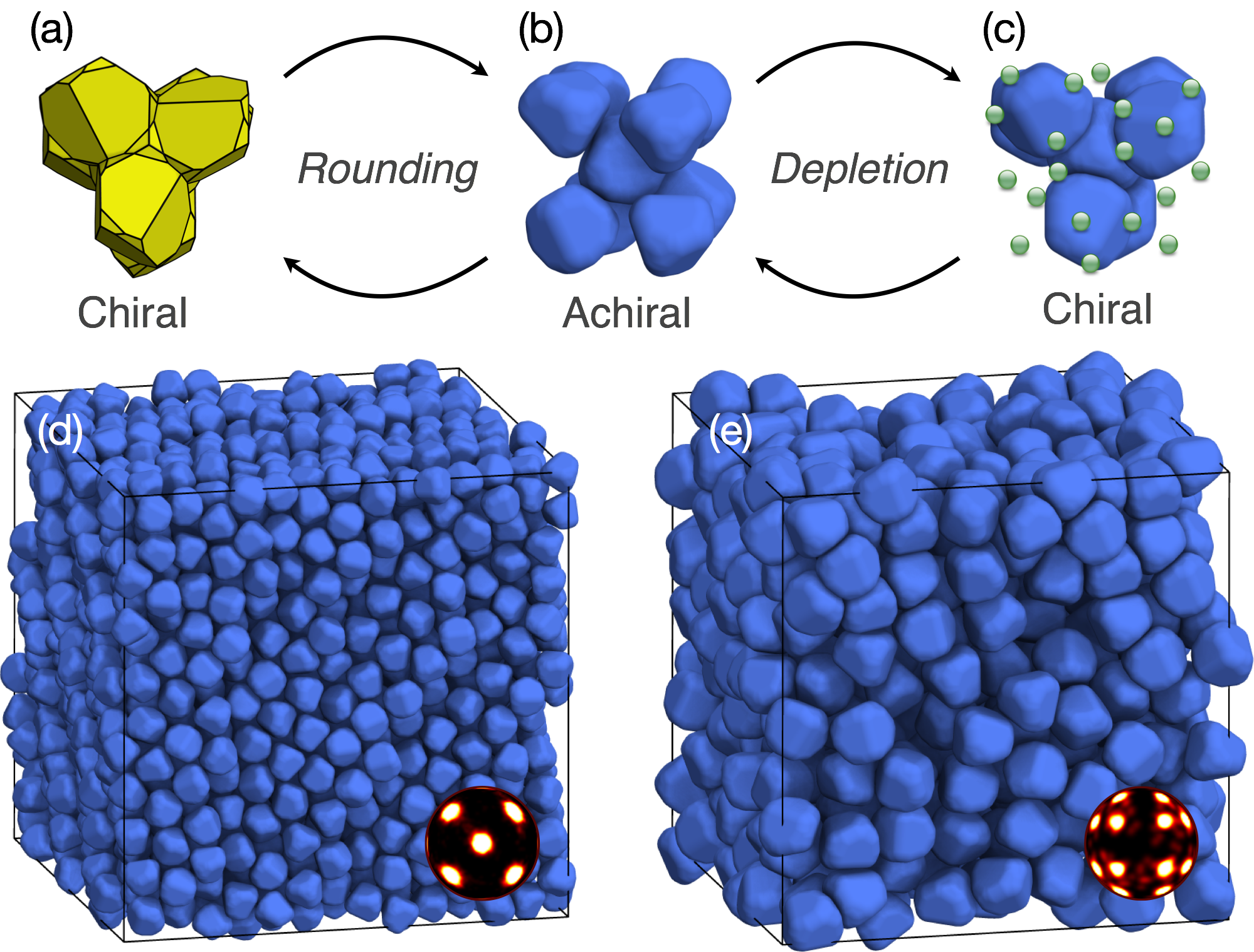}
\caption{
Reconfiguration between an achiral crystal and a homochiral crystal.
We compare the assembly of VPs of the SCC crystal (a)~using their original polyhedral shape, (b)~with rounded edges and corners, and (c)~additionally in the presence of depletants.
(d)~Rounding weakens DEFs and leads to the assembly of a BCC rotator crystal.
(e)~The effect of rounding can be reversed by depletion, which recovers SCC.
Depletants were simulated implicitly (see text for details).
They are shown as a guide in (c) and omitted in (e).
}
\label{fig3}
\end{figure}

\textit{Switching chirality.}---Can we systematically modify VP shape, and consequently the DEFs, to turn chirality off and on at will?
Recently it was shown that rounding (unrounding) polyhedrally shaped nanoparticles through the adsorption (desorption) of ligands can viably switch between two different crystals~\cite{Zhang2011}.
To test if this rounding approach can be used for chirality control, we modify the chiral VPs of the SCC crystal by means of vertex and edge rounding [Figs.~\ref{fig3}a,b].
Rounding is achieved in simulation by using the Minkowski sum of a polyhedron with a sphere.
Our results show that SCC crystals with the same handedness as the building blocks are assembled as the particles become increasingly rounded, but only up to IQ = 0.743~\cite{SM_Chiral}.
Above that value we observe an achiral BCC rotator crystal [Fig.~\ref{fig3}d].
That we do not find a gradual weakening of the preference for homochirality as for $\beta$-Mn, even when the particle sphericity is comparable to that of $\beta$-Mn VPs, can be understood by noticing that the low coordination number in SCC precludes the formation of a rotator SCC phase.
More about this trend will be discussed in the next section.

\begin{figure*}
\centering
\includegraphics[width=\textwidth]{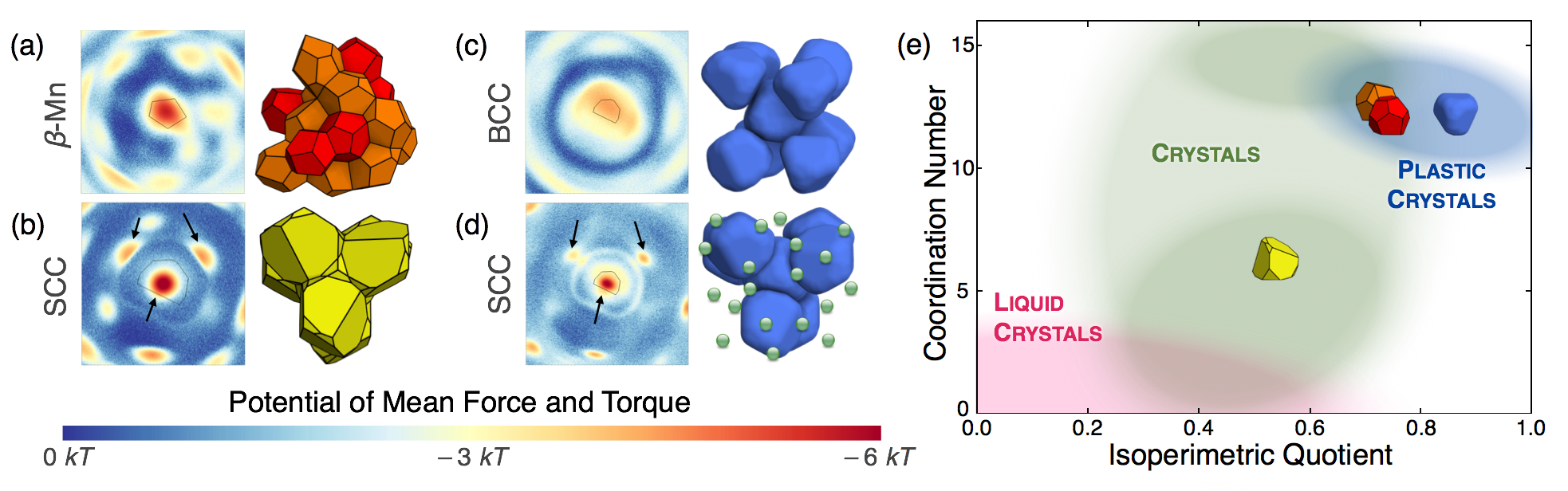}
\caption{
Quantifying the chiral strength of the particles used in this study.
(a-d)~PMFT diagrams for the four systems.
The diagrams were calculated by binning the frequency of finding two particle centers at a given separation vector, creating a histogram whose negative logarithmic value corresponds to the energy units reported.
The diagrams are two-dimensional slices taken through the first neighbor shell parallel to the largest facet on a shape.
They were calculated at the lowest densities at which each crystal was observed to self-assemble.
Color indicates the PMFT, from $0\,kT$ (dark blue) to $-6\,kT$ (dark red).
(e)~Coordination Number \textit{vs.} Isoperimetric Quotient plot derived from Ref.~\cite{Damasceno2012} including the shapes from (a-d).
}
\label{fig4}
\end{figure*}

Another approach to controlling the strength of DEFs for colloidal particles is \textit{via} the addition of depletants~\cite{vanAnders2014,Rossi2015}.
Non-adsorbing macromolecules or smaller colloids increase entropic forces between particles that are in close contact because it minimizes the excluded volume not accessible to the depletants~\cite{Lekkerkerker2011}.
Recently, depletion has been used to tune interparticle distances in colloidal crystals~\cite{Rossi2011b,Young2013}.
To test whether depletants can reverse the trend observed with rounding and re-assemble the homochiral SCC crystal, we use a recently developed grand canonical Monte Carlo algorithm that connects the hard colloid system to an external reservoir of penetrable hard sphere depletants~\cite{Glaser2015}.
The method accounts for depletion effects by sampling how a colloid trial move affects the free volume available to the depletants.
We combine colloids obtained from rounding SCC VPs with small depletants of radius $\sim0.1\sigma$ [Fig.~\ref{fig3}c], where $\sigma$ is the circumsphere radius of the colloids.
As our simulations demonstrate [Fig.~\ref{fig3}e], self-assembly in this system yields the original homochiral SCC crystal structure at depletant concentrations of $\geq20\%$ depletants in the reservoir.
This means the directionality of the depletion force, which is strongest at perfect face-to-face contact and thus highly directional, is able to undo the effect of rounding and restore homochirality.

\textit{Analysis of chirality strength.}---We quantify DEFs for the four systems presented above by determining potential of mean force and torque (PMFT) diagrams~\cite{vanAnders2014, vanAnders2014b}.
As is visible from the diagrams, $\beta$-Mn VPs have an entropic interaction of up to $5\,kT$ when faces are in contact [Fig.~\ref{fig4}(a)].
In contrast, the particles assembling homochiral crystals show stronger interaction of up to $6\,kT$ when big faces are in contact (central arrows in the diagrams) and up to $4\,kT$ when secondary faces are in contact (satellite arrows) [Fig.~\ref{fig4}(b,d)].
The rounded particle that assembles only BCC have a weaker interaction of about $4\,kT$ [Fig.~\ref{fig4}(c)].
In these examples, higher values for the PMFT help promote chirality by introducing directionality in the effective entropic coupling between neighbors.
Geometrical predictions from earlier work~\cite{Damasceno2012} are combined with new data in Fig.~\ref{fig4}(e).
They confirm once more that the particles in Fig.~\ref{fig4}(a,c) are too spherical to assemble a non-rotator phase, and thus too spherical to grow a homochiral crystal.

In summary, our findings suggest heuristics for the design of colloidal particles to self-assemble chiral entropic crystals with controlled handedness.
While a chiral space group can occur even for achiral building blocks, mirror-symmetry breaking of the colloid is a prerequisite to create an enantiomeric imbalance.
Sufficiently strong chiral forces are necessary for homochirality.
Such forces can be achieved via particle shape modifications alone, assuming the particle is not too spherical.
They can also be amplified by depletion effects.
Our results demonstrate a reconfigurable path for toggling between achiral and homochiral crystals via control of an external parameter, pushing the envelope towards the design of a new class of colloidal materials.

\acknowledgments
This work was supported partially by the U.\ S.\ Army Research Laboratory and the U.\ S.\ Army Research Office under grant number W911NF-15-1-0185. S.C.G.\ acknowledges support from DOD-ASD(R\&E) under Award No.\ N00244-09-1-0062.

\bibliography{chiralCrystal}

\end{document}